\documentclass[aps,prb,showpacs,preprintnumbers,psfrac,twocolumn]{revtex4}
\usepackage{amsmath}
\usepackage{epsfig}
\usepackage{graphicx}
\usepackage{dcolumn}
\usepackage{bm}

\newcommand{\la}{\langle}
\newcommand{\ra}{\rangle}

\def\gl{\lower.35em\hbox{$\stackrel{\textstyle>}{\textstyle<}$}}

\def\gapp{\lower.35em\hbox{$\stackrel{\textstyle>}{\sim}$}}
\def\lapp{\lower.35em\hbox{$\stackrel{\textstyle<}{\sim}$}}

\newcommand{\beq}{\begin{equation}} 
\newcommand{\eeq}{\end{equation}} 
\newcommand{\beqa}{\begin{eqnarray}} 
\newcommand{\eeqa}{\end{eqnarray}}

\newcommand{\om}{\omega}

\renewcommand{\=}{\!\!=\!\!}

\def\simleq{\; \raise0.3ex\hbox{$<$\kern-0.75em
      \raise-1.1ex\hbox{$\sim$}}\; }
\def\simgeq{\; \raise0.3ex\hbox{$>$\kern-0.75em
      \raise-1.1ex\hbox{$\sim$}}\; }

\def\la{{\langle}} 
\def\ra{{\rangle}}

\def\ni{\noindent}

\def\non{\nonumber }
\def\beq{\begin{equation} }
\def\eeq{\end{equation} }
\def\beqa{\begin{eqnarray}}
\def\eeqa{\end{eqnarray}}

\def\med{\frac{1}{2}}
\def\go{\vec g_1}
\def\gt{\vec g_2}

\begin{document}
\title{Symmetry-based  approach to electron-phonon interactions   in graphene}
\author{J. L. Ma\~nes}
\affiliation{Departamento de F\'{\i}sica  de la Materia Condensada\\
Universidad del Pa\'{\i}s Vasco,
Apdo. 644, E-48080 Bilbao, Spain}
\date{\today}

\begin{abstract}
We use the symmetries of monolayer graphene to write a set of constraints that must be satisfied 
by any electron-phonon interaction hamiltonian. The explicit solution as a series expansion in the momenta gives the most general, model-independent couplings between electrons and long wavelength acoustic and optical phonons. As an application, the possibility of describing elastic strains in terms of effective electromagnetic fields is considered in detail, with an emphasis on group theory conditions and the role of time reversal symmetry.

\end{abstract}

\pacs{63.20.Kr, 63.22.+m, 71.15.Rf} \maketitle


%






\section{Introduction}

Symmetry plays an obiquitous role in solid state physics, where invariance of the system under symmetry transformations imposes strong constraints on the form of hamiltonians and other physical observables.\cite{liu,brad} 
In computations  of phonon spectra, the  choice of convenient symmetry-adapted coordinates simplifies the diagonalization of the dynamical matrix.~\cite{mar} Instead of  a 
$3N\times 3N$ matrix, where $N$ is the number of atoms in the unit cell, we end up with a collection of 
smaller matrices, one per inequivalent IR (irreducible representation), with a dimension equal to the multiplicity of that representation in the `mechanical' representation. 
Similar  simplifications take place in  the computation of electronic bands.

An important {\it caveat} is that  the symmetry which is effective in simplifying the problem is not the whole space group  of the crystal, but  the  little group~\cite{liu} that preserves the wavevector of the  electrons or phonons.~\cite{foot2}
Thus, symmetry-based methods are  really powerful around points and lines of high symmetry in the Brillouin zone. Since electron-phonon interactions   involve generic wavevectors with trivial little groups, it would seem that  symmetry considerations can not be very helpful in that  case. 
Indeed, many computations of electron-phonon interactions are based on the use of specific models 
(electron-ion interactions,~\cite{zim,mah} modulated hoppings,~\cite{bar,piet,dress} phonon modulated electron-electron  interactions,~\cite{wood} etc), and little or no use is made of the symmetries of the system. 

On the other hand,   
long-wavelength phonons play a crucial role in many physical processes of interest, such as  transport phenomena, where electron-phonon interactions are essential.~\cite{zim,mah} The elements of the dynamical matrix for long wavelength phonons (acoustic and optical) can be expanded as a power series in the phonon wavevector $q$, and this expansion is strongly constrained by the little group of the $\Gamma$ point, which of course coincides 
with the   space group of the crystal. Unfortunately, low energy electrons lie near the Fermi surface, where  typical points lack any  symmetry.

Graphene~\cite{nov1,nov2} is a remarkable exception  in this regard. Instead of  an ordinary Fermi surface, undopped monolayer graphene has two isolated, high symmetry Fermi points. 
Thus, as long as we restrict ourselves to long wavelength phonons and low energy electrons, which are necesarily close to the Fermi points, the interaction hamiltonian  will be strongly constrained by symmetry. In fact, in Section~4 we will be able to write the most general interaction hamiltonian as an explicit  series expansion around the $\Gamma$ point (for phonon wavevectors) and the Fermi points (for electron wavevectors). 

One possible application of the hamiltonian obtained here is to  
many-body computations,~\cite{mah,wood2,neto} where  analytic expressions  are often preferred over numerically defined functions. In the usual approach   a concrete model, such as one based on phonon-modulated nearest neighbors hoppings, is used to compute an interaction hamiltonian 
which  later  is expanded  to the required order in the momenta.~\cite{wood,wood2,ando} Alternatively,  one can   take the  interaction hamiltonian  given in Section~4 as a power expansion, and keep  terms up to the order 
desired. Obviously,  less effort is involved by this method, and one is sure that all possible couplings  have been  included, whereas a particular model may miss some of then. The obvious drawback is that symmetry arguments by themselves do not give a clue as to  the values of the parameters. Another,  very different application of the electron-phonon hamiltonian is presented in Section~5, where we investigate a description of  elastic strains in terms of   effective electromagnetic fields.~\cite{foot1}

This paper is organized as follows. In Section~2 we review the symmetries of monolayer graphene and the transformation properties of electrons and phonons. Section~3 presents a study of the dynamical matrix including the form of the long wavelength normal modes.  The phonon-electron interactions are computed in Section~4, and the results are used in Section~5 to study the possibility of interpreting trilinear electron-phonon couplings in terms of effective scalar and vector potentials. The conclusions are presented in Section~6. 

\section{Symmetries}

We begin this section with a  description of electrons and phonons in 
monolayer graphene, including the choice  of  the symmetry-adapted modes to be  used in the rest of this paper. We then proceed to study their transformation properties.

\subsection{Electrons and phonons}
\begin{figure}[t]
\begin{center}
\includegraphics[angle=0,width=0.7\linewidth]{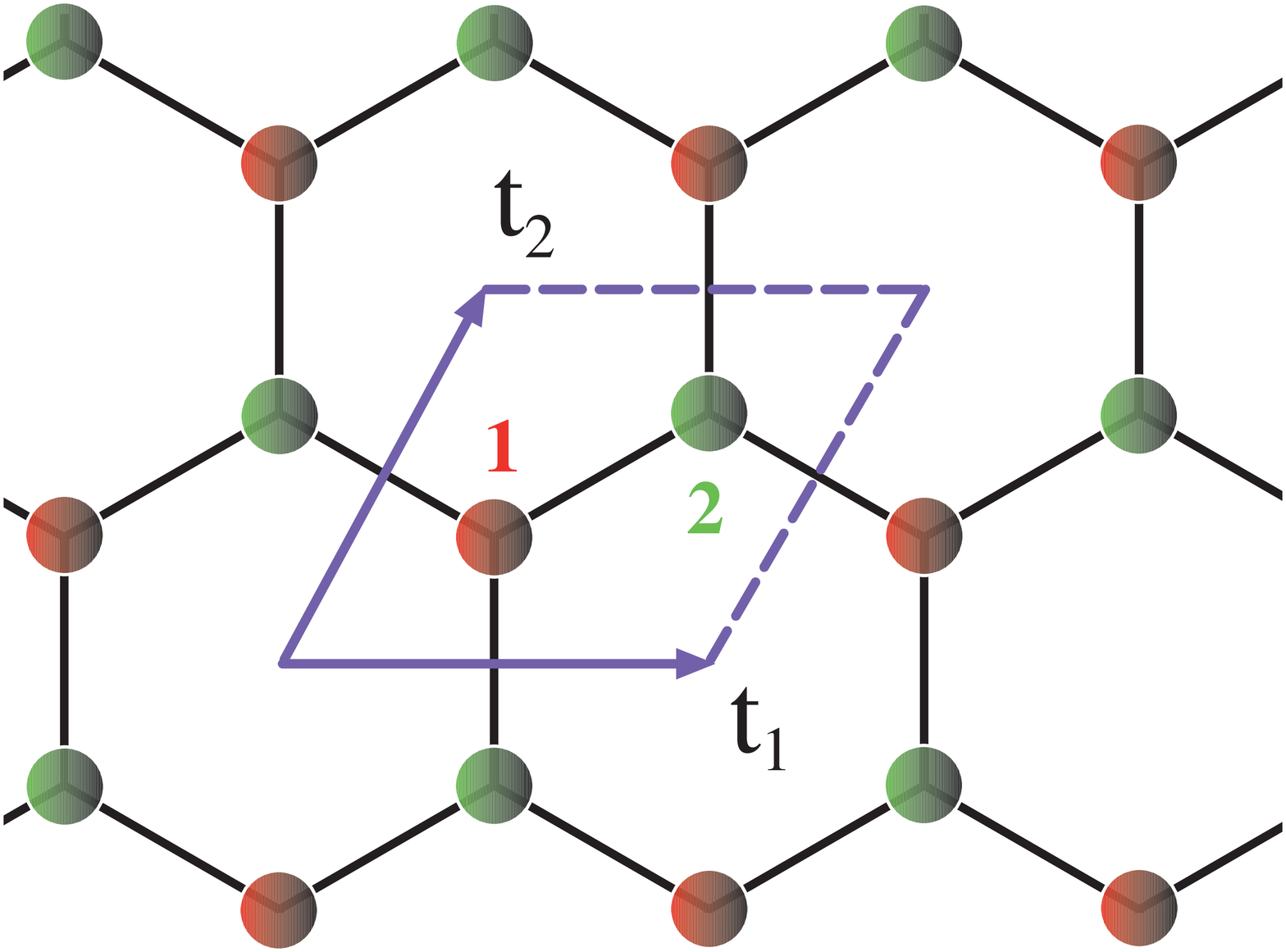}
\includegraphics[angle=0,width=0.5\linewidth]{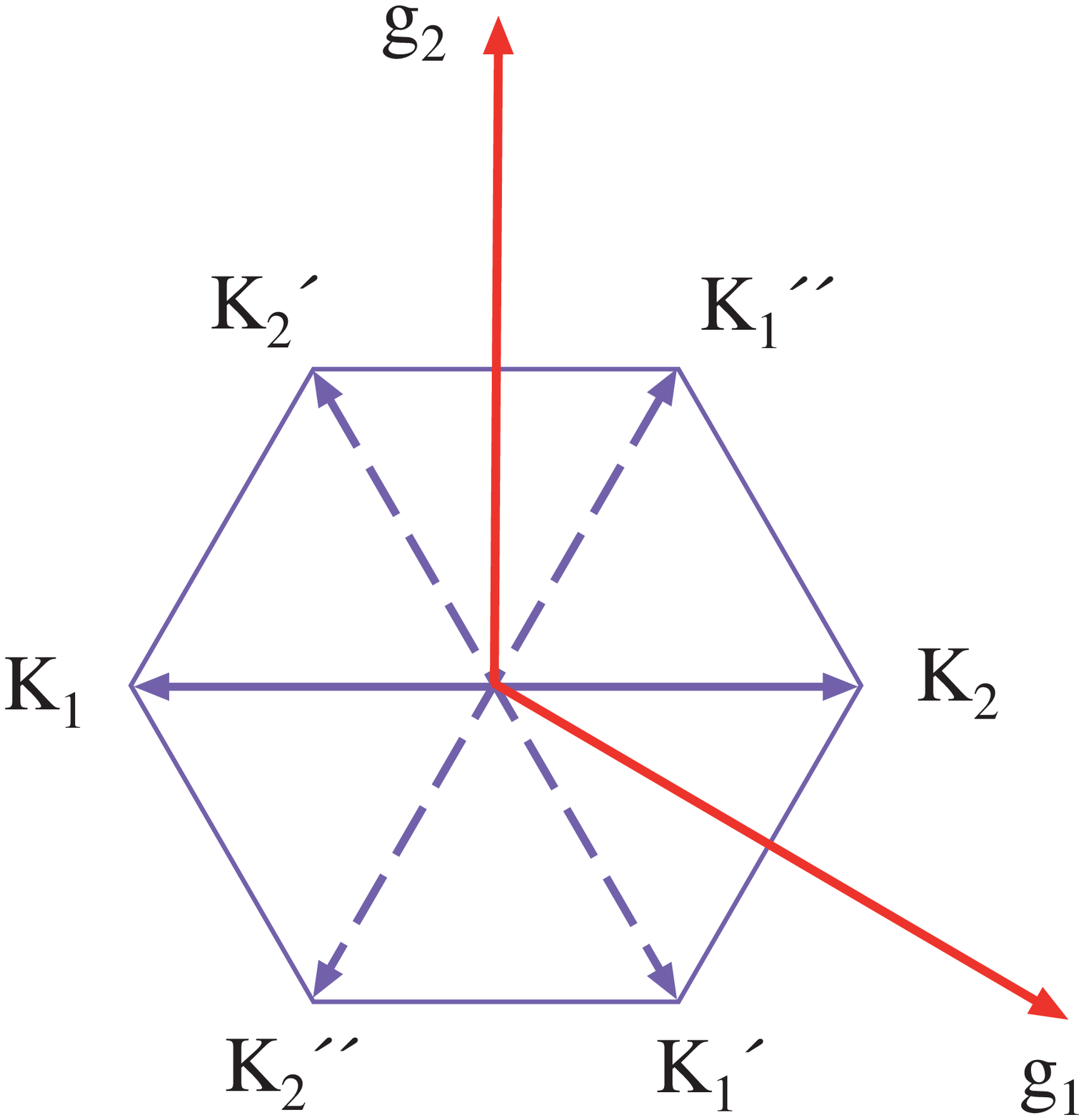}
\end{center}
\caption{ (Color online) Top: Direct lattice and unit cell for monolayer graphene. Bottom: First Brillouin zone and Fermi points. The vectors $\vec K\,'_1,\vec K\,''_1$ ($\vec K\,'_2,\vec K\,''_2$) are equivalent  to $\vec 
K_1$ ($\vec K_2$).}
\label{f1}
\end{figure}

Monolayer graphene  consists of a honeycomb      lattice. 
 The  corresponding  hexagonal  Bravais lattice  is generated by 
\hbox{$\vec t_1=(a \sqrt{3},0)$} and $\vec t_2=(a \sqrt{3}/2, 3a/2)$, where $a$ is the distance between nearest neighbours.  (See Fig.~\ref{f1}). The unit cell contains two atoms,  with positions  given by \hbox{$\vec r_1=\vec t_1/3+\vec t_2/3$} and $\vec r_2 = 2\vec r_1$. The reciprocal lattice is generated by $\vec g_1=
(2 \pi /\sqrt{3}a, -2 \pi/3a)$ and $\vec g_2=(0, 4 \pi/ 3 a)$, with $\vec t_i\cdot \vec g_j=2\pi \delta_{ij}$.
Corresponding  to  the two atoms in the unit cell,  we may define two Bloch wave functions
\beq\label{bloch}
\Phi_i (\vec K)=\sum_{\vec t} e^{i \vec K\cdot(\vec r_i+ \vec t)} \Phi (\vec r-\vec r_i-\vec t)
\; \;\; , \; \;\; i=1,2
\eeq
where the sum runs over all the points in the direct lattice, i.e., $\vec t = n_1 \vec t_1+n_2 \vec t_2$   
 and 
$\Phi(\vec r)$ is a real $\pi$-type atomic   orbital.   As is well known, a simple  tight-binding computation~\cite{wall,slon} yields  a spectrum with two Fermi points  located at $\vec K_1 = -2\go/3- \gt/3$ and $\vec K_2= -\vec K_1$. Near the two Fermi points, the hamiltonian can be linearized and  one finds
\beq\label{lin}
H(\vec K_1+\vec k)\sim{3\over 2}at\left( \begin{array}{cc} 0 & k^* \\ k & 0 \end{array} \right)\;,\;k\equiv k_x+ik_y
\eeq
where $t$ is the hopping integral, and $H(-\vec K_1+\vec k)=H^*(\vec K_1-\vec k)$. Thus, the low energy electronic excitations behave like massless Dirac fermions~\cite{gon1,gon2} with relativistic spectrum $E=\pm v_F |k|$. Henceforth we  assume our units are such that $v_F=3at/2=1$.

 In this paper we  want to consider the interaction between electrons near the Fermi points $\{\vec K_1,-\vec K_1\}$ and phonons $Q(\vec q)$ for small values of $q$.
 Due to invariance under reflection by the horizontal plane $\sigma_h$, only in-plane modes can couple linearly to electrons. Thus, for most of this paper we consider only in-plane  modes, which are specified by giving the horizontal displacements $u(\vec t\,)=(x_1,y_1,x_2,y_2)$ of the two atoms in the unit cell with origin at the lattice point $\vec t$.  See however Section~5, where quadratic couplings of out-of-plane acoustic modes are considered in the long wavelength limit.
  
 We can choose a basis of symmetry-adapted modes with particularly simple  transformation properties by noting that, under a point group transformation, a mode undergoes two changes: The vectors describing atomic displacements are rotated or reflected according to the vector representation, and there is a permutation among equivalent atoms. This last effect may give rise to annoying phases in the transformation matrices, which  can be avoided by incorporating appropriate phases into the definitions of the modes. The use of circular rather that linearly polarized modes simplifies the effect of rotations. 
 
 Our choice of in-plane modes  is given in Table~\ref{t1}. In the long wavelength limit $qa<\!<\!1$, the phases vanish $e_i(q)\!\to\! 1$ and the unprimed (primed) modes become  circularly polarized acoustic (optical) normal modes respectively (see Section~3 for details). 

\begin{table}[h]
\begin{tabular}{c | c c c c  }
 & $u_1$ & $u_2$ & $u'_1$ & $u'_2$\\
\hline \hline
$x_1$ & $e_1(q)$ & $e_1(q)$ &$e_1(q)$ & $e_1(q)$\\
$y_1$  & $i e_1(q)$ & $-i  e_1(q)$ & $i  e_1(q)$ & $-i e_1(q)$\\
$x_2$ & $e_2(q)$ & $e_2(q)$ & $-e_2(q)$ & $-e_2(q)$\\
$y_2$  & $ie_2(q)$ & $-ie_2(q)$ & $-ie_2(q)$ & $ie_2(q)$\\
\end{tabular}
\caption{ In-plane symmetry-adapted modes for monolayer graphene. Here  $e_i(q)=\med e^{i\vec q\cdot\vec r_i}$, where $\{\vec r_i\}$ are the equilibrium positions of the two atoms in the unit cell.}
\label{t1}
\end{table}
\ni This basis is  used to expand general in-plane displacements $u(\vec t\,)=(x_1,y_1,x_2,y_2)$ as follows
\beq\label{exp}
u(\vec t\,)=\sum_q\left(\sum_{i=1}^2 Q_i(\vec q\,) u_i(\vec q \,) +\sum_{i=1}^2 Q\,'_i(\vec q\,) u'_i(\vec q\,)\right) e^{i\vec q\cdot\vec t}
\eeq

\subsection{Transformation properties}

The space group   of monolayer graphene is symmorphic, with point group $D_{6h}$ (see ref.~\cite{brad} for conventions and notation).
$D_{6h}$ is generated by $ \{ C_6^+, \sigma_{v1}, \sigma_h \}$, where $ C_6^+$ is a counterclockwise rotation by $\pi/3$ around the $z$ axis, $\sigma_{v1}$ is a reflection by the $xz$ plane and $\sigma_h$
is a reflection by the horizontal $xy$ plane that contains the graphene layer. However, given the planar structure of the system and as long as  we are restricted to in-plane modes,  we will consider instead the normal subgroup $C_{6v}$, generated by $ \{ C_6^+, \sigma_{v1} \}$.
The little point group of $\vec K_1$ is $\hat G_{\vec K_1}=C_{3v}$, generated by $ \{ C_3^+, \sigma_{v1} \}$. The star of $\vec K_1$ consists of the two Fermi points $\{\vec K_1,-\vec K_1\}$.
Since  the binary rotation $C_2$ around the $z$ axis transforms $\vec K_1$ into -$\vec K_1$,  it is convenient to use $ \{ C_3^+, \sigma_{v1}, C_2 \}$ as a (redundant) set of generators for $C_{6v}$.

The action of the generators of $C_{6v}$ on the modes in Table~I is given in Appendix~A. As a result of a symmetry operation $g$, a  displacement with wavevector $q$ is transformed into a 
new displacement with wavevector $gq$ and coordinates $\{gQ_i, qQ'_i\}$. Defining $Q\equiv\{ Q_1,Q_2\}$ and  $Q'\equiv\{ Q'_1,Q'_2\}$, the transformation properties can be written
\beq\label{space}
gQ(gq)=T(g)\,Q(q)\;\;\; , \;\;\; gQ\,'(gq)=T\,'(g)\,Q\,'(q)
\eeq
where  $T'(C_2)=-T(C_2)=\mathbf 1$ is the unit matrix, and 
\beqa\label{trph}
T(\sigma_{v1})=-T'(\sigma_{v1})=\sigma_x=\left( \begin{array}{ccc}0 & 1\\ 1 &0 \end{array} \right)\non\\
 T(C^+_3)= T'(C^+_3)=\Omega\equiv\left( \begin{array}{ccc}\om^* & 0\\ 0 & \om \end{array} \right)
 \eeqa
with $\om\!\equiv\! \exp(2\pi i/3)$.

Another important symmetry is time reversal, which acts by complex conjugation. According to Table I, 
\beq
\mathcal{T}: u^*_1(q)=u_2(-q) \;\;\; , \;\;\; u^*_2(q)=u_1(-q) 
\eeq
with identical action on $u'_i$. Then, reality of a general displacement~(\ref{exp}) implies
\beq\label{tr}
Q^*(q)=\sigma_x Q(-q)  \;\;\; , \;\;\; Q'^*(q)=\sigma_x Q'(-q)
\eeq

We now turn our attention to the electrons. Since low energy electronic excitations are located near the two Fermi points, it is convenient to 
change variables, $\vec K =\pm \vec K_1 + \vec k$. Correspondingly, we define Bloch functions $\Phi_{i\pm}( k)$ in the neighbourhoods of the Fermi points 
\beq\label{phipm}
\Phi_{i\pm}( k)\equiv  \Phi_{i}(\pm\vec K_1+\vec k)
\eeq

The action of the generators of  $C_{6v}$ on the wavefunctions is given in Appendix~A. As long as we do not consider   the spin, 
time reversal acts simply by complex conjugation. According to~(\ref{bloch}), $\Phi^*_i(\vec K)=\Phi_i(-\vec K)$. Thus time reversal exchanges the two Fermi points and we have
\beqa
\mathcal{T}&:&   \Phi_{i+}(k)\to \Phi^*_{i+}(k)=\Phi_{i-}(-k)\non\\
&& \Phi_{i-}(k)\to \Phi^*_{i-}(k)=\Phi_{i+}(-k)
\eeqa

As both time reversal $\mathcal{T}$ and the binary rotation $C_2$ around the $z$ axis reverse the sign of the wavevector, 
 the combined symmetry $\mathcal{T} C_2$ preserves the Fermi points. Note however, that  the two atoms are exchanged
\beq\label{c2t}
\mathcal{T} C_2:   \Phi_{1+}(k)\leftrightarrow\Phi_{2+}(k)\;,\;  \Phi_{2-}(k)\leftrightarrow\Phi_{1-}(k)
\eeq
This combined symmetry  plays an important  role  in the computation of   electron-phonon interactions in Section~4 and is behind the topological stability of Fermi points in mono and multi-layer graphene.~\cite{exis}
\section{Dynamical Matrix}

The  modes in Table~I have simple transformation properties, but  they do not correspond to normal modes with well defined frequencies. In general, these must be obtained  by numerical diagonalization of the dynamical matrix.~\cite{mae,sed,orde}  However, in the long wavelength regime one can easily obtain useful analytical results by exploiting the symmetries of the system.

The symmetries constrain the form of the dynamical matrix $V(q)$ defined by
\beq
E_p= {m\over 2}\sum_q   \mathbf Q^\dagger(q) V(q)  \mathbf Q(q)
\eeq
where $\mathbf Q=\{Q,Q'\}$ stands for the four symmetry-adapted coordinates in~(\ref{exp}), $m$ is the mass of the carbon atom and the sum is over the first Brillouin zone. The hermitian matrix $V$ can be decomposed into $2\times 2$ blocks
\beq
 V(q)=\left( \begin{array}{ccc}D & M\\ M^\dagger & D' \end{array} \right)
\eeq
associated to the unprimed and primed modes. Then, invariance under a spatial symmetry~(\ref{space})  gives
\beq\label{inva}
\mathbf T(g)V(q)\mathbf T^\dagger(g)=V(gq)
\eeq
with
\beq
\mathbf T(g)=\left( \begin{array}{ccc}T(g) & 0\\ 0 &T'(g) \end{array} \right)
\eeq
whereas invariance under time reversal~(\ref{tr}) implies
\beq
\mathcal T V^*(q)\mathcal T=V(-q)\;\;,\;\;\mathcal T=\left( \begin{array}{ccc}\sigma_x & 0\\ 0 &\sigma_x \end{array} \right)
\eeq

Neglecting initially the acoustic-optical mixing due to $M$, we can obtain an approximation to acoustic frequencies and modes by solving a $2\times 2$ eigenvalue problem
\beq
D(q) u(q) = \om^2(q) u(q)
\eeq
where $u(q)$ is a linear combination  of the two unprimed modes given in table I. As shown in  Appendix~B, the symmetry constrains the form of the hermitian matrix $D$ to 
\beq\label{matD}
D(q)=\left( \begin{array}{ccc}a(q) & b(q) \\ b^*(q) & a(q) \end{array} \right)
\eeq
where $a(q)$ and $b(q)$ are even functions of $q$ that satisfy
\beqa\label{acous}
a(\om q)=a(q)&,& b(\om q)=\om b(q)\non\\ a(q^*)=a(q) &,& b(q^*)=b^*(q)
\eeqa
The eigenvalue problem is solved by 
\beq
\om_\pm^2(q)=a(q)\pm |b(q)|
\eeq
with normal modes given by
\beq
u_\pm (q)=
{1\over \sqrt  2}\bigl(u_1(q){b(q)\over |b(q)|}\pm u_2(q)\bigr)
\eeq

In the absence of acoustic-optical interactions, this is  the most general acoustic spectrum compatible with the symmetries of the system. For small $q$ this solution admits a simple geometric interpretation. Expanding  $a(q)$ and $b(q)$
\beqa\label{taylor}
a(q)&=&a_2|q|^2+a_4|q|^4+O(q^6)\non\\
b(q)&=&b_2q^{*2}+b_4 |q|^2q^{*2}+\tilde{b}_4 q^4+O(q^6)
\eeqa
where all the constants are real, and keeping only the quadratic terms yields
\beqa\label{lt}
\om^2_{L,T} (q)=(a_2\pm b_2) |q|^2
\eeqa
where the $+$ ($-$) sign goes with $\om^2_L$ ($\om^2_T$). The normal modes are
\beqa\label{lt}
u_L(q)&=&{1\over \sqrt 2|q|}\Bigl(q^*u_1(q)+q u_2(q)\Bigr) \non\\u_T(q)&=&-{i\over \sqrt 2|q|}\Bigl(q^*u_1(q)-q u_2(q)\Bigr)
\eeqa
These represent longitudinal (LA) and transverse acoustic (TA) modes respectively. To see this, note that for $qa<<1$, we have 
\beq
u_L(q)\simeq{1\over \sqrt 2} \left( \begin{array}{c}\cos \theta  \\  \sin\theta \\ \cos \theta  \\  \sin\theta \end{array} \right)
\; , \;
u_T(q)\simeq{1\over \sqrt 2} \left( \begin{array}{c}-\sin \theta  \\  \cos\theta \\ -\sin \theta  \\  \cos\theta \end{array} \right)
\eeq
where $q=|q| e^{i\theta}$, i.e. $\theta=\tan^{-1} (q_y/q_x)$. In the long wavelength limit, the normal modes  are completely independent of the coefficients in the dynamical matrix.

We can easily go beyond the   quadratic approximation in $q$  valid in the long wavelength limit by keeping more terms in the expansions~(\ref{taylor}).     Keeping just the first anisotropic contributions gives
\beq
\om_{L,T}^2(q)=(a_2\pm b_2) |q|^2 + |q|^4 (a_4\pm b_4\pm\tilde  b_4 \cos 6\theta)+O(q^6)
\eeq
One can  use perturbation theory to compute the leading corrections $u_{L,T}\to u_{L,T}+\delta u_{L,T}$  to the normal modes.
Taking as perturbation the fourth order corrections to $D$
\beq
\delta D=\left( \begin{array}{cc} a_4|q|^4 &b_4 |q|^2q^{*2}+\tilde{b}_4 q^4\\
b_4 |q|^2 q^{2}+\tilde{b}_4 q^{*4} & a_4|q|^4 \end{array} \right)
\eeq
yields
\beqa\label{ltm}
\delta u_L&\simeq&{u^\dagger_T\,\delta D\, u_L\over \om_L^2-\om_T^2} u_T\simeq -{\tilde b_4\over 2b_2}|q|^2\sin 6\theta\, u_T\non\\
\delta u_T&\simeq&{u^\dagger_L\,\delta D\, u_T\over \om_T^2-\om_L^2} u_L\simeq {\tilde b_4\over 2b_2}|q|^2\sin 6\theta\, u_L
\eeqa
Thus, beyond the long wavelength approximation  the normal modes are not purely transverse or longitudinal, and~(\ref{ltm}) shows that this longitudinal-transverse (L-T)   mixing is an $O(q^2)$ anisotropic  effect.

The solution to  the eigenvalue problem for the optical modes is very similar. The elements of the matrix 
$D'(q)$ satisfy the same constraints~(\ref{acous}), but the diagonal elements need not vanish for $q=0$. Keeping just the quadratic contributions
\beq
D'(q)=\left( \begin{array}{cc}\om_0^2+a'_2 |q|^2 & b'_2 q^{*2} \\ b'_2 q^{2} &\om_0^2+a'_2 |q|^2  \end{array} \right)
\eeq
we find that the optical normal modes $u'_L$ and $u'_T$ are given by~(\ref{lt}) in terms of  the primed modes $u_i'$, with normal frequencies
\beq
\om'^2_{L,T} (q)=\om^2_0+(a'_2\pm b'_2) |q|^2 
\eeq
Similarly,  L-T mixing for optical modes is still given by~(\ref{ltm}) if one uses primed constants.

We can also study acoustic-optical (A-O) mixing, which is induced by the matrix $M$. For example, keeping only the leading contributions to $M$ (see the Appendix~B for details.) 
\beq
M(q)\simeq  e_1\left( \begin{array}{ccc}0 & q \\ -q^* & 0 \end{array} \right)
\eeq
and using   perturbation theory  yields the following corrections to the LA and TA modes
\beqa\label{ao}
\delta u_L&\simeq&{i e_1\over \om_0^2}|q|(\sin 3\theta\, u'_L+\cos  3\theta\, u'_T)\non\\
\delta u_T&\simeq&{i e_1\over \om_0^2}|q|(\cos 3\theta\, u'_L-\sin  3\theta\, u'_T)
\eeqa
Thus A-O mixing is an $O(q/\om_0^2)$ anisotropic  effect. Exchanging primed and unprimed modes in~(\ref{ao}) gives the corrections to the optical modes. 
This mixing induces a common shift in the frequencies of the acoustic modes
\beq
\delta \om_L^2=\delta \om_T^2\simeq -{e_1^2\over \om_0^2}|q|^2
\eeq
while the opposite shift is induced in the optical frequencies.

\section{Electron-phonon couplings}
In this section we use a similar strategy to obtain the most general interaction hamiltonian compatible with the symmetries of the system. We consider an electron-phonon interaction hamiltonian of the form
\beqa\label{sec}
&&H_{e-ph}=\sum_{ij,k,q} H_{ij}(k,q,\mathbf Q) c^\dagger_{i+}(k+q/2)c_{j+}(k-q/2)\non\\
&+&\sum_{ij,k,q} H_{-ij}(k,q,\mathbf Q) c^\dagger_{i-}(k+q/2)c_{j-}(k-q/2)
\eeqa
where $c_{i\pm}(k)$ annihilates an electron in the state $\Phi_{i\pm}(k)$ given by~(\ref{phipm}) and  $\mathbf Q=\{Q,Q'\}$ stands for the four symmetry-adapted coordinates in~(\ref{exp}).  
The somewhat unusual parametrization of initial and final electron momenta \hbox{($k\mp q/2$)}  simplifies the form of the constraints.
The matrix element is given by
\beq\label{ion}
H_{ij}(k,q,\mathbf Q)=\la\Phi_{i+}(k+{q\over 2}) |V_{ei}|\Phi_{j+}(k-{q\over 2})\ra
\eeq
where $V_{ei}$ is the electron-ion potential.  
Since the two Fermi points are related  both  by  time-reversal and  by $C_2$, either symmetry   can be used to obtain the form of the hamiltonian near $\vec K_2=-\vec K_1$, giving
\beq\label{k2}
H_-(k,q,\mathbf Q)=H^*(-k,-q,\mathbf Q)
\eeq
where
\beq
H_{-ij}(k,q,\mathbf Q)\equiv\la\Phi_{i-}(k+q/2) |V_{ei}|\Phi_{j-}(k-q/2)\ra
\eeq
Thus, we can restrict ourselves to  the hamiltonian in the vicinity of $\vec K_1$.
 
On the other hand,  the combined symmetry $C_2 \mathcal T$~(\ref{c2t}) leaves the two Fermi points invariant and can be used to impose a constraint on  $H(k,q,C_2\mathbf Q)$
\beq\label{trev}
H(k,q,C_2\mathbf Q)=\sigma_x H^*(k,q,\mathbf Q)\sigma_x
\eeq
which is solved by
\beq\label{elecph}
H(k,q,\mathbf Q)=\left( \begin{array}{cc}\alpha(k,q,\mathbf Q) & \beta(k,q,\mathbf Q)  \\  \beta^*(k,q,C_2\mathbf Q) &\alpha^*(k,q,C_2 \mathbf Q) \end{array} \right)
\eeq
where $\alpha$ and $\beta$ are arbitrary complex functions.
Note that hermiticity of the hamiltonian does \textit{not} imply reality for $\alpha$. The point is that the matrix $H(k,q,C_2\mathbf Q)$ connects initial and final states with  different momenta, and instead we have
\beqa\label{herm}
H_{ij}(k,q,\mathbf Q)=\la\Phi_{i+}(k+{q\over 2}) |V_{ei}|\Phi_{j+}(k-{q\over 2})\ra\non\\=\la\Phi_{j+}(k-{q\over 2}) |V_{ei}|\Phi_{i+}(k+{q\over 2})\ra^*=
H^*_{ji}(k,-q,\mathbf Q)
\eeqa
which implies
\beqa\label{herm1}
\alpha(k,-q,\mathbf Q)&=&\alpha^*(k,q,\mathbf Q)\non\\
\beta(k,-q,\mathbf Q)&=&\beta(k,q,C_2\mathbf Q)
\eeqa

Finally, we must impose invariance  under the little point group $C_{3v}$
\beq\label{little}
T^\dagger(g) H(k,q,\mathbf Q) T(g)=H(gk,gq,g\mathbf Q)
\eeq
where $T(g)$ is given by~(\ref{trph}).
The hamiltonian (\ref{elecph}) together with the constraints (\ref{herm1},\ref{little}) contains the most general interactions beween electrons and in-plane phonons compatible space group  and time reversal symmetries. 
These  constraints are analyzed in detail in Appendix~C, where the general solution for the interaction hamiltonian is found in the form 
\beq\label{sum}
H(k,q,\mathbf Q)=H(k,q,Q)+H'(k,q,Q')
\eeq
with
\begin{widetext}
\beqa\label{master}
H(k,q,Q)&=&i\left( \begin{array}{cc}f(k,q) Q_1(q) - f(k,-q)^* Q_2(q) &\;\;  g_1(k,q)Q_1(q) +g_2(k,q)Q_2(q)  \\  & \\    g_2(k,q)^*Q_1(q) + g_1(k,q)^*Q_2(q)  &\;\; - f(k,-q) Q_1(q) +  f(k,q)^* Q_2(q) \end{array} \right)\non\\
&&\non\\
&&\non\\
H'(k,q,Q')&=&i\left( \begin{array}{cc}f'(k,q) Q'_1(q) - f'(k,-q)^* Q'_2(q) &\;\;  g'_1(k,q)Q'_1(q) +g'_2(k,q)Q'_2(q)  \\  & \\  -  g'_2(k,q)^*Q'_1(q) - g'_1(k,q)^*Q'_2(q)  &\;\;  f'(k,-q) Q'_1(q) -  f'(k,q)^* Q'_2(q) \end{array} \right)
\eeqa
\end{widetext}

The functions entering the hamiltonian must satisfy some simple conditions. In particular
the functions $f$, $f'$, $g_2$, and $g'_2$ must satisfy identical constraints, which we write only for $f$
\beq\label{fg2}
f(\om k,\om q)=\om f(k,q)\;\;,\;\; f(k^*,q^*)=f(k,q)^*
\eeq
whereas the conditions on $g_1$ and $g'_1$ are
\beq\label{g1}
g_1(\om k,\om q)= g_1(k,q)\;\;,\;\; g_1(k^*,q^*)=g_1(k,q)^*
\eeq
Besides, we must have
\beq\label{par}
g_i(k,-q)=-g_i(k,q)\;\;,\;\; g'_i(k,-q)=g'_i(k,q)
\eeq

Since $\om^3=1$, one immediately sees that any combination  of monomials of the form
\beq\label{mon}
k^n k^{*n'} q^m q^{*m'}
\eeq
with  real coefficients and $n-n'+m-m' =1\mod 3$ will satisfy~(\ref{fg2}). Imposing instead 
$n-n'+m-m' =0\mod 3$ we will get  solutions to~(\ref{g1}).  Finally, the conditions~(\ref{par})
 are satisfied by taking $m+m'$ odd (even)  for $g_i$ ($g'_i$).
 
 Eq.~(\ref{master}) together with the comments around eq.~(\ref{mon}) can be used to expand the interaction hamiltonian to any order in $q$ and $k$, giving  all possible in-plane phonon-electron couplings compatible with the symmetries of monolayer graphene, and are the main result in this paper. For the rest of this section we will study the long wavelength limit and some leading corrections.

Consider first the coupling to acoustic phonons.
Taking into account that the  interaction   should vanish for $q\to 0$, we can immediately write the leading contributions to the functions $f,g_1,g_2$ 
\beq\label{low}
f(k,q)\approx\sqrt   2\alpha_1q\; ,\; g_1(k,q)\approx 0 \; ,\; g_2(k,q)\approx \sqrt   2\beta_1q\
\eeq
where $\alpha_1$ and $\beta_1$ are real constants and the $\sqrt   2$ has been introduced for later convenience. This gives
\beqa\label{ham}
&&H(k,q,Q)\approx\\
&&i\sqrt   2\left( \begin{array}{cc}\alpha_1( q\, Q_1(q) + q^* Q_2(q)) & \,\beta_1 q\, Q_2(q) \\  \,\beta_1 q^*\, Q_1(q) &\alpha_1( q\, Q_1(q) + q^* Q_2(q)) \end{array} \right)\non
\eeqa

For some physical  applications it may be more convenient to express  the hamiltonian as a function of  normal  rather than symmetry-adapted coordinates. As we know from the previous section, the longitudinal and transverse modes in~(\ref{lt}) are   a good approximation to the actual normal modes in the long-wavelength limit. Using 
\beq
Q_1={q^*\over \sqrt  2|q|} (Q_L-iQ_T)\;\;\; , \;\;\; Q_2={q\over \sqrt  2|q|} (Q_L+iQ_T)
\eeq
the interaction hamiltonian takes the following form
\beqa\label{int}
&&H(k,q,Q)\approx\\
&&i|q|\left( \begin{array}{cc} 2\alpha_1 Q_L& \beta_1 e^{2i\theta}(Q_L+i Q_T)\\   \beta_1 e^{-2i\theta}(Q_L-i Q_T)  & 2\alpha_1 Q_L\end{array} \right) \non
\eeqa
where $\theta=\tan^{-1}(q_y/q_x)$ is the phase of $q\!=\!|q|e^{i\theta}$. 
This shows that TA phonons do not couple 
diagonally to  electrons
 in the long-wavelength limit,  which is parametrized by the  
two real coulplings $\alpha_1$ and $\beta_1$. This form of the hamiltonian can be used to write~(\ref{sec})  in terms of phonon creation and annihilation operators,~\cite{mah} with
\hbox{$Q_{L,T}(q)=(2 \om_{L,T})^{-1/2}(a_{L,T}(q)+a^\dagger_{L,T}(-q))$}.

 Note that even if one is primarily interested  in acoustic phonons, because of  the acoustic-optical mixing discussed in the last section, one may have to consider the couplings of electrons to  optical modes as well.   Since the couplings to optical modes do not have to vanish as $q\to 0$, the leading contributions to the hamiltonian are independent of $q$. These are  found to be
\beqa\label{opt}
&&H(k,q,Q')\approx
i\sqrt 2  \beta'_1\left( \begin{array}{cc}0& Q'_1\\ -Q'_2  &0\end{array} \right)\non\\
&=&i \beta'_1 \left( \begin{array}{cc}0&e^{-i\theta} (Q'_L-iQ'_T)\\  -e^{i\theta} (Q'_L+iQ'_T)  &0\end{array} \right)
\eeqa
where $\beta'_1$ is a real constant. 

We can easily  include higher powers of the momenta.  For instance, for acoustic phonons the contribution of order $O(q^2)$   is parametrized by a single real constant  $\alpha_2$  and is purely diagonal and proportional to the Pauli matrix $\sigma_z$
\beq
\Delta H(k,q,Q)=\alpha_2 |q|^2(Q_L\sin 3\theta+Q_T \cos 3\theta )\sigma_z
\eeq
This is qualitatively different from the leading contribution~(\ref{int}) due to the anisotropy in the strength of the electron-phonon couplings and the diagonal contribution of TA phonons.
The contribution of order $O(q)$ for optical phonons is  proportional to the unit matrix 
\beq
\Delta H'(k,q,Q')=i\alpha'_2 |q|Q'_L\mathbf 1
\eeq
and there is also an $O(k)$ term given by
\beq
\Delta H'(k,q,Q')=
\alpha'_3 |k|(Q'_L\sin(\varphi-\theta)+Q'_T\cos(\varphi-\theta) ) \mathbf 1
\eeq
where $\varphi=\tan^{-1}(k_y/k_x)$ .
The contributions of order $O(qk)$ are more complicated and depend on five real constants. 
Note that as we move away from the $\Gamma$-point the LA and TA modes given by~(\ref{lt}) are no longer a good approximation to the normal modes of the system, and one has to take into account  the L-T and A-O mixings  considered in the previous section.

Electron-phonon interactions can be  rewritten in terms of electrons and holes.  The eigenstates for the electron hamiltonian~(\ref{lin}) are given by
\beq\label{eigen}
\Phi_\pm(k)={1\over\sqrt{2}}\Bigl( e^{-i\varphi/2}\Phi_1(k)\pm e^{i\varphi/2}\Phi_2(k)\Bigr)
\eeq
and this can be used to obtain the matrix elements  between electron eigenstates, with $H_{++}$ and $H_{--}$ ($H_{+-}$ and $H_{-+}$) corresponding to intraband (interband ) transitions. Of course, if we decide to  include higher powers of $k$ in the electron-phonon hamiltonian, for consistency we must  also go beyond the low-energy Dirac hamiltonian~(\ref{lin}). The corrections to the electronic hamiltonian   are fixed by symmetry. Note, in particular, that the electronic  hamiltonian can be considered as the $\mathbf Q$-independent part of~(\ref{ion}). Then~(\ref{elecph}) reduces to 
\beq\label{nlin}
H(k)=\left( \begin{array}{cc}\alpha_0(k) & \beta_0(k)  \\  \beta_0^*(k) &\alpha_0^*(k) \end{array} \right)
\eeq
where hermiticity~(\ref{herm1}) implies that $\alpha_0$ is real. The little group constraints~(\ref{little2}) simplify to 
\beqa
C_3^+&:& \alpha_0(\om k)=\alpha_0(k)\;\;,\;\; \beta_0(\om k)=\om^* \beta_0(k)\non\\
\sigma_{v1} &:&  \alpha_0(k^*)=\alpha_0(k)^*\;\;,\;\; \beta_0(k^*)= \beta_0(k)^*
\eeqa
which can be easily solved to any order in $k$. Taking the chemical potential at half filling as the origin of energies and $v_F=1$, the first few terms are
\beqa
 \alpha_0(k)&=& \alpha_{0,2}|k|^2+\alpha_{0,3}(k^3+k^{*3})+\ldots\non\\
  \beta_0(k)&=&k^*+\beta_{0,2} k^2+\beta_{0,3} |k|^2 k^*+\ldots
 \eeqa
 where all the constants are real. The eigenstates are still given by~(\ref{eigen}) if one replaces the phase of $k$ by the phase of  $\beta^*_0(k)$.
\section{Elastic strains as effective electromagnetic fields}

Here we show that the couplings of long wavelength phonons to electrons have some similarities ---and differences--- 
with those of the scalar and vector  potentials for the electromagnetic field. We show, in particular, that both in-plane and out-of-plane strains can mimic some of  the effects of electric and magnetic fields.

The minimal coupling prescription $k_i\to k_i+A_i$ on~(\ref{lin}) gives
\beqa\label{min}
H(k)&=&\left( \begin{array}{cc} \Phi & k^*+A^* \\ k+A &\Phi \end{array} \right)\non\\
H_-(k)&=&\left( \begin{array}{cc}\Phi & -k-A \\ -k^*-A^*&\Phi \end{array}\right)
\eeqa 
where $A=A_x+i A_y$ is the vector potential in complex notation and $\Phi$ is the scalar (electric) potential.
For the purposes of this section it is convenient to change from  the circularly polarized  modes of table I to   linear modes $u_x$, $u_y$
\beqa\label{xy}
u_1={1\over\sqrt 2}(u_x + i u_y)\;& ,&\;  u_2={1\over\sqrt 2}(u_x - i u_y)\Rightarrow\non\\
Q_1={1\over\sqrt 2}(Q_x - i Q_y)\;& ,&\; Q_2={1\over\sqrt 2}(Q_x + i Q_y)
\eeqa
Then, the acoustic hamiltonian~(\ref{ham}) can be rewritten by using 
\beqa
i \sqrt 2(q Q_1+q^* Q_2)=2 i (q_x Q_x + q_y Q_y)\non\\
i \sqrt 2 q Q_2=2 i (q_x Q_x -q_y Q_y)-2(q_xQ_y+q_y Q_x)\non\\
i \sqrt 2 q^*  Q_1=2 i (q_x Q_x -q_y Q_y)+2(q_xQ_y+q_y Q_x)
\eeqa
This agrees  with the hamiltonian  given in ref.~\cite{ando} where the off-diagonal elements (second and third lines)
were obtained by expanding a phonon-modulated hopping, while the magnitude of the  diagonal element, known as   `deformation potential',  was  estimated in a nearly free electron model.
Our derivation shows that both diagonal and off-diagonal terms are uniquely determined by symmetry.

Taking Fourier transforms with $iq_i\to \partial_i$ and comparing with~(\ref{min}) finally yields
\beqa\label{eff}
\Phi(K_1)=\Phi(K_2)=2\alpha_1(\partial_x Q_x+\partial_y Q_y)\non\\
A_x(K_1)=-A_x(K_2)=2\beta_1(\partial_x Q_x-\partial_y Q_y)\non\\
A_y(K_1)=-A_y(K_2)=-2\beta_1(\partial_x Q_y+\partial_y Q_x)
\eeqa
where we have used~(\ref{k2}) to obtain the electron-phonon hamiltonian around $\vec K_2$.
These effective fields can be written in terms of elastic strains with the usual definitions
\beq
u_{xx}=\partial_x Q_x\;,\;u_{yy}=\partial_y Q_y\;,\; 2u_{xy}=\partial_x Q_y+\partial_y Q_x
\eeq
Thus we have the remarkable result that, around each Fermi point, static strains can mimic the effects of external electric and magnetic fields.

We also see that the couplings  of electrons   near the two Fermi points are identical for   the scalar potential, but
differ by a  sign for the vector potential. Thus, for static strains they will experience the same 
 `electrostatic' field, but opposite `magnetic' fields given by
 \beqa
 B_z(K_1)&=&-B_z(K_2)=\partial_x A_y(K_1)-\partial_y A_x(K_1)\non\\
 &=&-2\beta_1\left[(\partial^2_x-\partial^2_y)Q_y+2\partial_x\partial_y Q_x\right]\non\\
 &=&-2\beta_1\left[2\partial_x u_{xy}+\partial_y(u_{xx}-u_{yy})\right]
 \eeqa

We now  address  the following question: Is the (partial) identification between 
elastic strains and effective electromagnetic fields peculiar to  graphene alone, or can we expect to find it in other systems? 
The first observation is straightforward:  The minimal coupling prescription $\vec k\to \vec k+\vec A$ gives rise to the required type of interactions for the vector potential only if the electron system satisfies the Dirac equation. An ordinary non-relativistic equation would produce terms of the type $\vec k \cdot\vec A$, where $\vec k$ is the electron momentum, unlike the leading  phonon couplings~(\ref{ham}), which are $k$-independent.

But even with  Dirac points, further conditions have to be satisfied. 
 The effective fields~(\ref{eff}) associated to  acoustic phonons are of the form $\partial_i  Q_j$, where both $\partial_i$ and $Q_i$ belong to the vector representation $V$. The vector potential also belongs to $V$. In the continuum, $V\times V$ and $V$ have opposite parities, and $\partial_i  Q_j$ and $A$ can not transform  equivalently. Thus, the possibility of describing elastic strains as effective fields is a lattice effect. More concretely, we need a lattice without inversion symmetry.
 
 But even on a lattice this identification can be partial at best. The reason is that the vector potential is odd under time reversal, whereas phonons and their derivatives are even. Since time reversal takes $\vec K$ to $-\vec K$, as long as $-\vec K$ is not equivalent to $\vec K$,  we can still hope for a partial  identification valid around individual Fermi points. This will be 
 possible only if the representations $V$ and $V\times V$ of the \emph{little} group $G_{\vec K}$ have 
 at least one  irreducible representation in common, i.e., if $V^3$ contains the identity (or trivial) representation.

We now discuss how these conditions are met by monolaye  graphene.  The littele group is $\hat G_{\vec K_1}=C_{3v}$ with a vector  representation that decomposes according to~\cite{brad}
\beq\label{vector}
V=A_1(z)+E(x,y)
\eeq
For in-plane modes only $E$ is relevant, and the use of elementary group-theory techniques~\cite{liu,brad} yields 
\beqa\label{EE}
E\times E&=& A_1(\partial_x Q_x+\partial_y Q_y)+A_2(\partial_x Q_y-\partial_y Q_x)\non\\
&+& E(\partial_yQ_y-\partial_x Q_x, \partial_x Q_y+\partial_y Q_x)
\eeqa
Here we recognize the L.H.S. of (\ref{eff}) as the basis for the irreducible representations $A_1$ and $E$. But the fact that  
 the vector potential $(A_x,A_y)$  and certain components of the strain tensor transform equivalently under the little point group $C_{3v}$ is not sufficient to  guarantee
 that elastic strains can mimic a magnetic field around each Fermi point, for they could still couple to the two atoms in the unit cell with different signs. By~(\ref{c2t}), the two atoms are exchaged under 
 $C_2\mathcal T$. Now, the vector potential is odd under $C_2$ and $\mathcal T$, whereas the 
 the in-plane components of the elastic strain are even under both symmetries, and the two minus signs cancell each other. However, electrons at the two Fermi points see effective vector potentials  which differ by the sign. On the other hand, 
  the scalar potential $\Phi$ belongs to the trivial representation
$A_1$ and is even under time reversal. As a consequence, the effective scalar potential takes the same sign on $\vec K_1$ and $\vec K_2=-\vec K_1$.

This analysis can be  extended to include the effects of so-called ripples or long wavelength 
deformations perpendicular to the grapehene sheet. Such  ripples  make the two dimensional graphene sheet thermodinamically stable and  have been   recently observed in individually suspended  sheets~\cite{nat}. Although we have restricted ourselves to in-plane modes, acoustic out-of-plane modes 
can be easily incorporated in the long wavelength limit (see Section~6 for the possibility of a more general analysis): 

According to~(\ref{vector}) a vertical displacement $Q_z$ belongs to the IR $A_1$ of $C_{3v}$. On the other hand,  $Q_z$ is odd under reflection by the horizontal plane $\sigma_h$, whereas electronic wavefunction bilinears are necessarily even. This forces us to consider \textit{quadratic} functions of $Q_z$. By translation invariance along the $z$ axis only derivatives of 
$Q_z$ are acceptable, and we have to consider quadratic functions of $\partial_x Q_z$ and $\partial_y Q_z$. The basis for the IRs $A_1$ and $E$ are now given by
\beqa
A_1\left( (\partial_x Q_z)^2+(\partial_y Q_z)^2\right)\non\\
E\left((\partial_y Q_z)^2-(\partial_x Q_z)^2, 2\partial_x Q_z \partial_y Q_z \right)
\eeqa
The discussion after Eq.~(\ref{EE}) applies also in this case, implying that  the couplings  of electrons   near the two Fermi points are identical for   the scalar potential, but
differ by a  sign for the vector potential. Thus, the associated effective fields are given by
\beqa\label{eff2}
\Phi(K_1)=\Phi(K_2)=\gamma\left( (\partial_x Q_z)^2+(\partial_y Q_z)^2\right)\non\\
A_x(K_1)=-A_x(K_2)=\delta\left((\partial_y Q_z)^2-(\partial_x Q_z)^2\right)\non\\
A_y(K_1)=-A_y(K_2)=2\delta\partial_x Q_z \partial_y Q_z
\eeqa
where $\gamma$ and $\delta$ are real coupling constants.
 
 We close this section by noting  that frozen optical modes can also give rise to effective magnetic fields.\cite{eff} Indeed, the substitution of~(\ref{xy}) in~(\ref{opt}) gives
\beqa
H(k,q,Q')&=&i\sqrt 2  \beta'_1\left( \begin{array}{cc}0& Q'_1\\ -Q'_2  &0\end{array} \right)\non\\
 &=& \beta'_1 \left( \begin{array}{cc}0&Q'_y+iQ'_x\\  Q'_y-iQ'_x\ &0\end{array} \right)
\eeqa
Comparing with the minimally coupled hamiltonians~(\ref{min}) yields the identifications
\beqa
A'_x(K_1)&=&-A'_x(K_2)=\beta'_1  Q'_y\non\\
A'_y(K_1)&=&-A'_y(K_2)=-\beta'_1 Q'_x
\eeqa
and the magnetic fields
\beq
B'_z(K_1)=-B'_z(K_2)=-\beta'_1 (\partial_x Q'_x+\partial_y Q'_y)
\eeq 

\section{Discussion}

In this paper we have exploited the fact that, for many physical processes of interest in monolayer graphene,  the wavevectors of electrons and phonons  lie near points of high symmetry in the Brillouin zone. Even though  the little groups for non-vanishing $k$ and $q$ are generically  trivial, the proximity to  points of high symmetry impose strong constraints on the series expansions of observables around them. This is  analogous to the situation  in Landau's theory~\cite{liu} of second order phase transitions, where the dependence on the order parameter of observables  in the low symmetry phase is determined by the space group of the 
{\it high} symmetry phase. Here, $q$ and $k$ play the role of order parameters, `breaking' the symmetries of the $\Gamma$ and Fermi points respectively.
As a consequence, the method presented in this paper  may be useful in other systems with Fermi points, such as multilayer graphene.~\cite{exis,nil1,nil2,nov} Semimetals with  small electron and hole pockets around high symmetry points are also good candidates.

Note that, in spite of the fact that we use tight-binding Bloch functions as our starting point, our results are 
model-independent. The reason is that only the symmetry properties of the wavefunctions are used, and 
the general electronic hamiltonian~(\ref{nlin}) describes  any doublet of states transforming according to the small representation $E$ of the little group.

The relatively involved constraints~(\ref{herm1},\ref{little}) on the electron-phonon hamiltonian are reduced to extremely simple conditions on the few functions which appear in the general solution~(\ref{master}) in Section~4 ---so simple, indeed, that they can be solved explicitly. In the process the general functions $\alpha$ and $\beta$ are written as {\it linear} combinations of the symmetry-adapted coordinates. By dropping the linearity assumption we could extend our approach to  multiphonon processes. One can easily check that the constraints~(\ref{herm1},\ref{little}) are still valid for nonlinear functions depending on several modes and momenta,  with
\beq
\alpha(k,q,\mathbf Q)\to\alpha\bigl(k,\{ q_l \},\{ \mathbf Q_l ( q_l )\}\bigl)
\eeq
and other obvious replacements. Note that at the non-linear level out-of-plane modes have to be included for consistency, and the whole symmmetry $D_{6h}$ rather than $C_{6v}$ has to be used. The analysis becomes more involved and will be the object of future work.

Inspired by  a mechanism
first proposed in ref.~\cite{kosh} for the case of dislocations in multivalley conductors, effective magnetic fields induced by elastic distorsions have been recently suggested as a way to explain the strong suppression of weak localization in graphene.~\cite{morpu,strong} This differs from our results in Section~5 in that they consider   effective gauge fields which  are {\it quadratic} in the out-of-plane modes. Our analysis in the previous section shows that in-plane modes give rise to efffective fields which are linear in the strains. For physical applications of the treatment in Section~5 to dislocations and other deffects see ref.~\cite{eff} 

Our method can be easily extended to accomodate external fields. For instance, in the presence of an electric field along the $z$ axis, out-of-plane phonons will couple linearly to the electrons. This may be an externally applied electric field, or  the effect of a substrate which breaks the symmetry under reflections on the sample plane. The recently observed ripples in suspended graphene sheets~\cite{nat}  
also break the reflection symmetry $\sigma_h$ and can play the role of a  `background field',  inducing new  electron-phonon interactions.


\section*{Acknowledgments}

It is a pleasure to thank F. Guinea for motivating this work  and for  very  useful discussions and suggestions, particularly in reference to  Section~5. This work has been    supported in part by the Spanish Science Ministry under Grant FPA2005-04823.

\appendix

\section{Transformations of modes and wavefunctions}

The action of the generators of $C_{6v}$ on the electron and phonon wavevectors is given by the vector representation. This is simpler in  complex notation 
\beq
C^+_3 q= \om q \;\;\; , \;\;\; \sigma_{v1} q=q^*\;\;\; , \;\;\; C_2 q=-q \;\;\; , \;\;\;  \om\equiv e^{2\pi i/3}
\eeq
where  $q=q_x+iq_y$. 

Careful inspection of Table~\ref{t1} shows that the  generator $C_3^+$  acts identically on primed and unprimed symmetry-adapted modes, namely 
\beq
C_3^+: u_1(q)\to \om^*u_1(\om q)\;\;\; , \;\;\;  u_2(q)\to \om u_2(\om q)
\eeq
where $\om\equiv\exp(2\pi i/3)$.
The actions of $\sigma_{v1} $ and $C_2$, however, are different for the two types of  modes
\beqa
\sigma_{v1} &: &u_1(q)\to u_2( q^*)\;\;\; ,  \;\;\; u_2(q)\to u_1(q^*)\non\\
& &u'_1(q)\to -u'_2( q^*)\;\;\; ,  \;\;\; u'_2(q)\to -u'_1(q^*)\non\\
C_2&: &u_i(q)\to - u_i(- q)\;\;\; , \;\;\;  u'_i(q)\to  u'_i(- q)
\eeqa
Correspondingly, the symmetry-adapted coordinates $Q_i$ and $Q'_i$ in~(\ref{exp}) have  transformation properties
\beq
gQ(gq)=T(g)\,Q(q)\;\;\; , \;\;\; gQ\,'(gq)=T\,'(g)\,Q\,'(q)
\eeq
where  $T'(C_2)=-T(C_2)=\mathbf 1$ is the unit matrix, and 
\beqa
T(\sigma_{v1})=-T'(\sigma_{v1})=\sigma_x=\left( \begin{array}{ccc}0 & 1\\ 1 &0 \end{array} \right)\non\\
 T(C^+_3)= T'(C^+_3)=\Omega\equiv\left( \begin{array}{ccc}\om^* & 0\\ 0 & \om \end{array} \right)
 \eeqa
 Note that at the $\Gamma$-point the unprimed (primed) modes transform according to the two dimensional representation $E_1$ ($E_2$) of $C_{6v}$.

 The action of the generators of the little group $C_{3v}$ on the electronic wavefunctions is given by 

\beqa\label{trans}
C^+_3&:&\Phi_{1+}(k)\to \om \Phi_{1+}(\om k)\; ,\;\Phi_{2+}(k)\to \om^* \Phi_{2+}(\om k)\non\\
&&\Phi_{1-}(k)\to \om^* \Phi_{1-}(\om k)\; ,\;\Phi_{2-}(k)\to \om \Phi_{2-}(\om k)\non\\
\sigma_{v1}&:& \Phi_{1+}(k)\leftrightarrow \Phi_{2+}(k^*) \; ,\;  \Phi_{1-}(k)\leftrightarrow\Phi_{2-}(k^*)
\eeqa
The binary axis $C_2$ connects the two Fermi points
\beq\label{trans2}
C_2: \Phi_{1+}(k)\leftrightarrow \Phi_{2-}(-k)\; ,\; \Phi_{2+}(k)\leftrightarrow \Phi_{1-}(-k)
\eeq
Equations~(\ref{trans},\ref{trans2}) imply that, at the Fermi points \hbox{($k=0$)}, the  wavefunctions $\{\Phi_{i\pm}(0)\}$ form the basis for  the  $4$-dimensional IR $E$ of the space group,~\cite{foot3} 
with star $\{\vec K_1,-\vec K_1\}$.

\section{Symmetries of the dynamical matrix}

Since according to~(\ref{inva}) invariance under the generators  of $C_{3v}$ impose identical  constraints on $D$ and $D'$, we write these only for $D$
\beq\label{c3}
C_3^+:  D(\om q)=\Omega D(q) \Omega^* \; , \; \sigma_{v1}: D(q^*)=\sigma_x D(q) \sigma_x
\eeq
The nondiagonal block $M$ satisfies the same constraint under $C_3^+$, but gets an additional minus sign under $  \sigma_{v1}$
\beq\label{sig}
\sigma_{v1}: M(q^*)=-\sigma_x M(q) \sigma_x
\eeq
Invariance under $C_2$ implies that $D$ and $D'$ are even functions of $q$, whereas $M$ is  odd 
\beq\label{c2}
C_2: D(-q)=D(q)\; , \;  D'(-q)=D'(q) \; , \;  M(-q)=-M(q)
\eeq

Finally, time-reversal invariance~(\ref{tr}) imposes the additional constraint
\beq
 D(-q)=\sigma_x D^*(q)\sigma_x 
\eeq
which is also satisfied by $D'$ and $M$. Combining this equation with~(\ref{c3})-(\ref{c2}) gives
\beq
D(q^*)=D^*(q) \; , \;   D'(q^*)=D'^*(q) \; , \;  M(q^*)=M^*(q)
\eeq
These constraints are solved by 
\beq
D(q)=\left( \begin{array}{ccc}a(q) & b(q) \\ b^*(q) & a(q) \end{array} \right)
\eeq
where $a(q)$ and $b(q)$ are even functions of $q$ that satisfy
\beqa
a(\om q)=a(q)&,& b(\om q)=\om b(q)\non\\ a(q^*)=a(q) &,& b(q^*)=b^*(q)
\eeqa
with identical solution for $D'(q)$. For the off-diagonal block the solution is 
\beq
M(q)=\left( \begin{array}{ccc}d(q) & e(q) \\ -e(q^*) & -d(q^*) \end{array} \right)
\eeq
where  $d(q)$ and $e(q)$ are odd functions of $q$ that satisfy the following constraints
\beqa
d(\om q)=d(q) \;\;\;&,& \;\;\; e(\om q)=\om e(q)\non\\
 d(q^*)=d^*(q) \;\;\; &, &\;\;\; e(q^*)=e^*(q)
\eeqa
These imply the expansions 
\beqa
d(q)&=& d_3 q^3+ \tilde d_3 q^{*3}+O(q^5)\non\\
e(q)&=& e_1 q+ e_3 |q|^2 q+O(q^5)
\eeqa
where all the constants are real.

\section{Little group constraints on the electron-phonon hamiltonian}

Substitution of~(\ref{elecph}) into~(\ref{little}) yields the following constraints on  
the complex functions $\alpha$ and $\beta$ 
\beqa\label{little2}
C_3^+&:& \alpha(\om k,\om q,C_3^+\mathbf Q)=\alpha(k,q,\mathbf Q)\non\\\ &&\beta(\om k,\om q,C_3^+\mathbf Q)=\om^*\beta(k,q,\mathbf Q)\non\\
\sigma_{v1}&:& \alpha(k^*,q^*,\sigma_{v1}\mathbf Q)=\alpha^*(k,q,C_2\mathbf Q)\non\\\ &&
\beta(k^*,q^*,\sigma_{v1}\mathbf Q)=\beta^*(k,q,C_2\mathbf Q)
\eeqa
where the argument of $\mathbf Q$ is always the second one  in the function, i.e., $\alpha(\om k,\om q,C_3^+\mathbf Q)$ actually stands for $\alpha(\om k,\om q,C_3^+\mathbf Q(\om q))$.
The general solution to the constraints can be found by writing
\beqa\label{alph}
\alpha(k,q,\mathbf Q)&=&\alpha(k,q,Q)+\alpha'(k,q,Q')\non\\ 
\beta(k,q,\mathbf Q)&=&\beta(k,q,Q)+\beta'(k,q,Q')  
\eeqa
with
\beqa
\alpha(k,q,Q)&=&i\big(f_1(k,q) Q_1(q)+f_2(k,q) Q_2(q)\big)\non\\
\beta(k,q,Q)&=&i\big(g_1(k,q) Q_1(q)+g_2(k,q) Q_2(q)\big)
\eeqa
and equivalent expressions for $\alpha'$ and $\beta'$  in terms of primed variables. Then, substitution of $\alpha$  into the hermiticity conditions~(\ref{herm1}) gives
\beq
f_2(k,q)=-f_1(k,-q)^*
\eeq
which implies
\beq
\alpha(k,q,Q)=i\big(f(k,q) Q_1(q)-f^*(k,-q) Q_2(q)\big)
\eeq
with identical results for $\alpha'$. Imposing the conditions~(\ref{herm1}) on $\beta$ and  $\beta'$ gives
\beq
g_i(k,-q)=-g_i(k,q)\;\;,\;\; g'_i(k,-q)=g'_i(k,q)
\eeq

Substitution   into the little group constraints~(\ref{little2})   shows that the functions $f$, $f'$, $g_2$, and $g'_2$ must satisfy identical conditions, which we write only for $f$
\beq
f(\om k,\om q)=\om f(k,q)\;\;,\;\; f(k^*,q^*)=f(k,q)^*
\eeq
whereas the conditions on $g_1$ and $g'_1$ are
\beq
g_1(\om k,\om q)= g_1(k,q)\;\;,\;\; g_1(k^*,q^*)=g_1(k,q)^*
\eeq
The general solution for the electron-phonon hamiltonian is then given by~(\ref{sum},\ref{master}).



\begin{thebibliography}{99}
\bibitem{foot2}{Symmetry elements that do not belong to the little group can be used to relate  observables at different points of the Brillouin zone, but they do not impose constraints on the form of  observables at a particular point.}
\bibitem{foot3}{As  the space group is symmorphic, the IRs at the $K$ point can be labelled by those of the little point group $C_{3v}$, which  are $A_1$, $A_2$ and $E$. Throughout this paper we follow   the notation and conventions of ref.~\cite{brad}}
\end{thebibliography}
\end{document}